\documentstyle[12pt]{article}
\def\bold#1{\setbox0=\hbox{$#1$}%
      \kern-.025em\copy0\kern-\wd0
      \kern.05em\copy0\kern-\wd0
      \kern-.025em\raise.0433em\box0 }
\def\eea{\end{eqnarray}}
\def\bea{\begin{eqnarray}}
\def\eeas{\end{eqnarray*}}
\def\beas{\begin{eqnarray*}}
\def\ee{\end{equation}}
\def\be{\begin{equation}}
\def\bdm{\begin{displaymath}}
\def\edm{\end{displaymath}}

\def\Tr{\mbox{Tr}}
\def\skp{\epsilon}

\renewcommand{\thefootnote}{\fnsymbol{footnote}}
\begin{document}
\begin{titlepage}

\begin{center}

\vspace*{2.0cm}
\hfill TAN-FNT-97-04
\vspace*{2.0cm}

{\large\bf $\Big|\Delta S\Big| = 1$ HADRONIC WEAK DECAYS OF HYPERONS IN A SOLITON MODEL}
\vskip 1.5cm

{Norberto N. SCOCCOLA\footnote[2]{Fellow
of the CONICET, Argentina.}}
\vskip .2cm
{\it
Physics Department, Comisi\'on Nacional de Energ\'{\i}a At\'omica,
          Av.Libertador 8250, (1429) Buenos Aires, Argentina.\\}

\vskip 2.cm
September 1997
\vskip 2.cm
{\bf ABSTRACT}\\
\begin{quotation}
We study the parity violating hyperon non-leptonic weak decays in the  
three flavor Skyrme model. We follow the approach in which the symmetry 
breaking terms in the action are diagonalized exactly within
the collective coordinate approximation. We show that although this 
method introduces some configuration mixing, the $\Delta I =1/2$ rule
is numerically well satisfied. In addition, and in contrast to previous 
calculations,
we show that not only the relative amplitudes are in good agreement with
the empirical values but also their absolute values. The issue of whether
the strong interaction enhancement factors should be included in soliton
calculations is also addressed.
\end{quotation}
\end{center}
\end{titlepage}

\renewcommand{\thefootnote}{\arabic{footnote}}

Nonleptonic weak decays are still one of the least understood aspects
of low energy weak interactions. The main difficulty is related with
the evaluation of hadronic matrix elements of the weak hamiltonian.
In the absence of good hadronic wave functions obtained directly 
from QCD one has to resort to effective low energy models. In this
sense, quark models with QCD enhancement factors have been quite
successful in predicting hyperon S-wave decay 
amplitudes (see Ref.\cite{DGH86} and references therein).
The situation in soliton models seemed to be rather
different, however. Calculations performed in the mid-eighties 
showed\cite{PT85,TF86} that although octet dominance was present
in such models (that is, $\Delta I = 1/2$ rule was well
satisfied) and predicted relative amplitudes were in good agreement
with the empirical values, their absolute values turned out to
be far too small. Such calculations
have been done using the so-called ``perturbative" approach
to the $SU(3)$ Skyrme model. In such an approach, $SU(3)$ collective
coordinates are introduced to quantize the soliton and 
symmetry breaking terms are treated in first order perturbation 
theory. It is well-known by now that this naive approach leads to 
very poor predictions even for the hyperon spectra\cite{Pra85}. 
Moreover, in those calculations the pion decay constant $f_\pi$ 
(taken as a free parameter) has to be adjusted to less than one half 
of its empirical value in order to reproduce some of the observed
mass splittings. This small value of $f_\pi$ was believed to be
at the origin of the failure in reproducing the absolute weak 
decay amplitudes. With the introduction of more refined methods
to treat chiral symmetry breaking terms the situation was somewhat
improved. In Ref.\cite{KSO90} it was shown that within a framework
in which hyperons are treated as soliton-kaon bound systems
\cite{CK85} the calculated matrix elements are indeed larger
than those obtained in the perturbative approach. However, they
still fall quite below the empirical ones. In this paper we will
show that the correct absolute values can be naturally obtained within
a scheme in which $SU(3)$ collective coordinates are used but
symmetry breaking terms are diagonalized exactly. This approach
was pioneered by Yabu and Ando\cite{YA88} and improved by several
authors (for a review see Ref.\cite{Wei96}). As a result of this
diagonalization process, configuration mixing appears. One might
wonder whether this fact, together with the inclusion of kinematic 
symmetry breaking terms (needed to obtain good predictions for different
observables) will not induce deviations from the empirical
well satisfied $\Delta = I =1/2$ rule. As we will see this is
not the case for a reasonable parameterization of the model.

As well-known\cite{MRR69}, using PCAC and isospin symmetry the seven 
different hyperon non-leptonic amplitudes can be expressed in terms of five 
independent ones. They are related to the parity
conserving weak hamiltonian according to
\footnote{Here and in what follows we use the phase convention given in
Ref.\cite{DGH86}. Note also that $f_\pi$ is defined in such a way that
empirically $f_\pi= 93 \ MeV$.}
\begin{eqnarray}
A(\Lambda^0_-) \! \! & \! \! = \! \! & \! \!  
- {1\over{\sqrt{2} f_\pi}} < n | H_{w,\Delta S = 1}^{pc} | \Lambda > 
\: \: \: \: ,  \: \: \: \:
A(\Sigma^+_0) = {1\over{2 f_\pi}} < p | H_{w,\Delta S = 1}^{pc} | \Sigma^+ > \nonumber \\
A(\Xi^-_-) \! \! & \! \! = \! \! & \! \!
 - {1\over{\sqrt{2} f_\pi}} < \Lambda | H_{w,\Delta S = 1}^{pc} | \Xi^0 > 
\: \: \: \: , \: \: \: \:
A(\Sigma^-_-) = {1\over{f_\pi}} < n | H_{w,\Delta S = 1}^{pc} | \Sigma^- > \\
A(\Sigma^+_+) \! \! & \! \! = \! \! & \! \!
 - {1\over{\sqrt{2} f_\pi}} 
\left[ < p | H_{w,\Delta S = 1}^{pc}  | \Sigma^+ > +
\sqrt{2} \ < n | H_{w,\Delta S = 1}^{pc}  | \Sigma^0 > 
\right] \ .\nonumber
\label{indampl}
\end{eqnarray}
For the parity conserving weak interaction hamiltonian we use the
Cabbibo current-current form
\begin{equation}
H_{w,\Delta S = 1}^{pc} = \tilde G
\ J^{L}_{\mu,\pi^-} \ J^{L,\mu}_{K^+} \ .
\end{equation}
Here, $J^{L}_{\mu,a}$ are the left hadronic currents and 
$\tilde G = G_F \sin \theta_c \cos \theta_c / \sqrt2$, where
$G_F$ is the Fermi coupling constant and $\theta_c$ is the Cabbibo
angle. Moreover, we have used
the shorthand notation ${\pi^-} = 1 - i 2$ and ${K^+} = 4 + i 5$.
Within the
Skyrme model the currents $J^{L}_{\mu,a}$ can be obtained as Noether currents 
of the effective
chiral action supplemented with appropriate symmetry breaking terms. 
We use the form
\be
\Gamma = \Gamma_{SK} + \Gamma_{WZ} + \Gamma_{SB} \ ,
\label{action}
\ee
where $\Gamma_{SK}$ is the Skyrme action
\be
\Gamma_{SK} =
\int d^4 x \Big\{ {f^2_\pi \over 4} 
\Tr\left[ \partial_\mu U (\partial^\mu U)^\dagger \right]
+
 {1\over{32 \skp^2}}
 \Tr\left[ [U^\dagger \partial_\mu U , U^\dagger \partial_\nu U]^2\right] \Big\} \, .
\ee
Here, $\epsilon$ 
is the dimensionless Skyrme parameter. Furthermore the chiral field 
$U$ is the non--linear realization of the pseudoscalar octet. 
$\Gamma_{WZ}$ is the Wess-Zumino action :
\bea
\Gamma_{WZ} &=& - {i N_c \over{240 \pi^2}}
\int d^5x \epsilon^{\mu\nu\rho\sigma\tau}
\Tr[ L_\mu L_\nu L_\rho L_\sigma L_\tau] 
\eea
where $L_\mu = U^\dagger \partial_\mu U$ and $N_c=3$ is the number of colors.
Finally, $\Gamma_{SB}$ represents the symmetry breaking terms:
\bea
\Gamma_{SB} & = &\int d^4x \left\{
 { f_\pi^2 m_\pi^2 + 2 f_K^2 m_K^2 \over{12} }
 \Tr \left[ U + U^\dagger - 2 \right] 
%\right. \nonumber \\
%& & \qquad \left.
+ \sqrt{3}  { f_\pi^2 m_\pi^2 - f_K^2 m_K^2 \over{6} }
\Tr \left[ \lambda_8 \left( U + U^\dagger \right) \right] \right.
\nonumber \\
& & \qquad
 \left.
+ { f_K^2 - f_\pi^2\over{12} }
\Tr \left[ (1- \sqrt{3} \lambda_8)
\left(
U (\partial_\mu U)^\dagger \partial^\mu U + 
U^\dagger \partial_\mu U (\partial^\mu U)^\dagger \right)
\right] \right\} \, .
\label{sb}
\eea
\noindent
Here $f_K$ is the kaon decay constant while $m_\pi$ and $m_K$ are the 
pion and kaon masses, respectively. 

A straightforward calculation shows that the corresponding left current 
can be expressed as
\begin{eqnarray}
J^{L}_{\mu,a} &=& - {i \ f_\pi^2\over 2} \ Tr \Big( \lambda_a \ R_\mu \Big)
                  + {i\over{8 e^2}} \ 
                  Tr \Big( [ \lambda_a , R_\nu ] [ R_\mu,R^\nu] \Big) 
           \nonumber \\
              & & + {N_c\over{48 \pi^2}} \epsilon^{\mu \nu \alpha \beta}
          Tr \Big( \lambda_a \ R_\nu R_\alpha R_\beta \Big) -
          i { f_K^2 - f_\pi^2 \over{12}} 
       \ Tr \Big( (1-\sqrt{3} \ \lambda_8) [ U, \lambda_a ] R_\mu \Big) \ ,   
\end{eqnarray}
where $R_\mu = \partial_\mu U  U^\dagger$. The contribution of the different terms 
in Eq.(\ref{action}) to the left current can be easily recognized.

In the soliton picture we are using the strong interaction 
properties of the low--lying $\frac{1}{2}^+$ and $\frac{3}{2}^+$ 
baryons are computed following the standard $SU(3)$ collective 
coordinate approach to the Skyrme model. We introduce the 
{\it ansatz} 
\begin{equation}
U({\bf r}, t) = A(t) \ \left( 
\begin{array}{cc} 
c + i \mbox{\boldmath $\tau$} \cdot 
{\hat{\mbox{\boldmath $r$}}} \ s & 0 \\
0  & 1 
\end{array}
\right) 
\ A^\dagger(t) \ 
\label{ansatz}
\end{equation}
for the chiral field.
Here we have employed the abbreviations $c= \cos F(r)$ and 
$s=\sin F(r)$ where $F(r)$ is the chiral angle which parameterizes 
the soliton. The collective rotation matrix $A(t)$ is $SU(3)$ valued. 
Substituting the configuration Eq.(\ref{ansatz}) into $\Gamma$
yields (upon canonical quantization of $A$) the collective Hamiltonian.
Its eigenfunctions and eigenvalues are identified as the baryon 
wavefunctions $\Psi_B(A)=\langle B |A\rangle$ and masses $m_B$.
Due the symmetry breaking terms in $\Gamma_{sb}$ this Hamiltonian 
is obviously not $SU(3)$ symmetric. As shown by Yabu and Ando \cite{YA88}
it can, however, be diagonalized exactly. This diagonalization 
essentially amounts to admixtures of states from higher dimensional 
$SU(3)$ representations into the octet ($J=\frac{1}{2}$) and decouplet 
($J=\frac{3}{2}$) states. This procedure has proven to be quite successful 
in describing the hyperon spectrum and static properties \cite{Wei96}. 

Using the ansatz Eq.(\ref{ansatz}) in the expression of the left current
we obtain that, to leading order in $N_c$, the weak hamiltonian can be written as
\begin{eqnarray}
H_{w,\Delta S = 1}^{pc} &=& - \phi_{SK} \ R_{{\pi^-},a} R_{{K^+},a} +
               \phi_{WZ} \ R_{{\pi^-},8} R_{{K^+},8} \nonumber \\
            & & \qquad 
              - \phi_{SB} \ \left[ \left( {1 + 2 R_{8,8} \over2} \right) 
                            R_{{\pi^-},a} R_{{K^+},a} +  
                            \left( {2 + R_{8,8} \over3} \right)
                            R_{{\pi^-},8} R_{{K^+},8} \right] ,
\label{red-weak}
\end{eqnarray}
where
\begin{eqnarray}
\phi_{SK} &=& { {\tilde G} f_\pi^4 \over3} \int d^3r 
\Bigg[ \left(F'^2 + 2 {s^2\over{r^2}} \right) + {4\over{e^2 f_\pi^2}}
{s^2\over{r^2}} \left( 2 F'^2 + {s^2\over{r^2}} \right) \nonumber \\
& & \qquad \qquad \qquad \qquad \qquad +{2\over{e^4 f_\pi^4}}   
{s^2\over{r^2}} \left( F'^4 + 4 F'^2 {s^2\over{r^2}} + {s^4\over{r^4}} \right)
\Bigg] \ , \\[3.mm]
\phi_{WZ} &=& {{\tilde G} N_c^2\over{48 \pi^4}} \int d^3r \ F'^2 {s^4\over{r^4}} \ ,
\\[3.mm]
\phi_{SB} &=& { {\tilde G} f_\pi^2 \over9}  \left( f_K^2 - f_\pi^2 \right) 
\int d^3r \ (1-c) \left[ \left(F'^2 + 2 {s^2\over{r^2}} \right) + 
{4\over{e^2 f_\pi^2}} {s^2\over{r^2}} \left( 2 F'^2 + {s^2\over{r^2}} \right)
\right] \ .
\end{eqnarray}
The $SU(3)$ rotation matrices are defined by
\begin{equation}
R_{a,b} = {1\over2} Tr \Big( \lambda_a A^\dagger \lambda_b A \Big) \ .
\end{equation}

For simplicity, in Eq.(\ref{red-weak}) we have not written the contribution quadratic
in $(f_K^2 - f_\pi^2)$ since for empirical values of the decay constants it turns
out to be numerically completely negligible.

In the present model the hyperon decay amplitudes can be computed by taking the
matrix elements of the hamiltonian Eq.(\ref{red-weak}) between the hadronic states
expressed as linear combinations of SU(3) D-functions. For this purpose
it is convenient to use the Clebsch-Gordan decomposition of the collective operators
appearing in the weak hamiltonian. One obtains
\begin{eqnarray}
R_{{\pi^-},a} R_{{K^+},a} &=& - {3\sqrt{6}\over{5}} D^{8}_{{1\over2},0} - 
{1\over{10}} D^{27}_{{1\over2},0}
                            - {1\over{\sqrt{20}}} D^{27}_{{3\over2},0} 
\label{ope1}\\
 \! \! \! \! \! \!  R_{8,8} R_{{\pi^-},a} R_{{K^+},a} 
\! \! & \! \! =  \! \! &  \! \!
 - \sqrt{2\over{75}} D^{8}_{{1\over2},0} - {29\over{35}} D^{27}_{{1\over2},0}
                          -  {2\over{7\sqrt{30}}} D^{64}_{{1\over2},0} +
      {1\over{28\sqrt{5}}}  D^{27}_{{3\over2},0} - {\sqrt{3}\over{21}}  D^{64}_{{3\over2},0} 
\label{ope2}
\end{eqnarray}
and similar relations for those containing $R_{{\pi^-},8} R_{{K^+},8}$.
Here, the left lower index of the SU(3) D-functions $I = {1\over2}, {3\over2}$ stands for
$(Y,I,I_3) = (1,I,-{1\over2})$ while the right lower index for $(0,0,0)$.

At this stage we note the potential advantages and drawbacks of the present
approach with respect to the perturbative calculations of
Refs.\cite{PT85,TF86}. On one hand the use of an exact diagonalization allows for
the use of empirical meson decay constants \cite{Wei96}. This will certainly lead 
to an improvement of the decay amplitudes absolute values. On the other hand,
since as a consequence of this diagonalization baryon wavefunctions
contain higher $SU(3)$ representations the relevant 
matrix elements of the collective operators Eqs.(\ref{ope1},\ref{ope2}) will
not be, in general, ``octet dominated". In this sense, it is not clear
whether the $\Delta I = 1/2$ rule will be well satisfied as it was the case in
the perturbative calculation. 

We turn now to the numerical calculations. We take the meson masses to their
empirical values $m_\pi = 138 \ MeV$ and $m_K = 495 \ MeV$. 
Moreover, we use the empirical value $f_\pi = 93 \ MeV$.
As already stressed
several times in the literature the use of $f_\pi \ne f_K$ is essential to reproduce
the observed mass differences of the low lying octet and decouplet baryons.
Therefore, we take $f_K = \ 120 MeV$ which together with $\skp = 4.10$ 
gives a very good overall description of various hyperon properties
\cite{Wei96}. As well-known with these parameters the soliton mass turns out to be, 
at tree level, quite large as compared 
to the value needed to reproduce the empirical nucleon mass. However, in the last
few years it was shown \cite{Mou93} that, within the $SU(2)$ soliton model,
the inclusion of one-loop meson corrections reduces that value significantly.
Very recently \cite{Wal97} this same conclusion was extended to the $SU(3)$ 
models. Therefore, at present time, the parameter set above can be considered
as the optimal one within the approach adopted here. 

Our results for the decay amplitudes are given in Tables~1 and 2. In Table~1 we
show the decay amplitude taken with respect to $A(\Lambda^0_-)$ while in Table~2
we give the absolute value of this particular amplitude. The results are presented in 
this way to make easier the comparison with the values obtained in other models.
In fact, also shown in Table~1 are those of the perturbative approach (PTA) to the
SU(3) soliton model \cite{PT85}, the bound state soliton model (BSA) \cite{KSO90}
and the empirical values taken from Ref.\cite{DGH86}. The value for the quark 
model (QM) that appears in Table~2 has been taken from Ref.\cite{TT81}.
Note that in Table ~1 only the values of the independent amplitudes Eq.(\ref{indampl}) 
are given. The reason
is that all the corresponding model calculations (and the QM as well) 
are based on the use of PCAC and isospin symmetry which implies
\begin{equation}
{A(\Lambda^0_0)\over{A(\Lambda^0_-)}} = {A(\Xi^0_0)\over{A(\Xi^-_-)}} = 
- {1\over{\sqrt{2}}} \ .
\end{equation}
Although the $\Lambda$-ratio is not known empirically, the $\Xi$-ratio is 
${A(\Xi^0_0)\over{A(\Xi^-_-)}}\Big|^{emp} = - 0.75$. This is usually taken
as an indication that PCAC and isospin symmetry can be used in this framework.

In Table~1 we observe that the relative values of the decay amplitudes
are quite well reproduced in our model. Of particular interest is
$A(\Sigma^+_+)$. In the limit in which the $\Delta I = 1/2$ rule is exactly
satisfied this amplitude is zero. We
see that our value, although small, does not vanish. In fact, it nicely
reproduces the small departure from the $\Delta I = 1/2$ rule verified by the empirical
amplitudes. The reason for the smallness of our calculated value even in the
presence of configuration mixing is twofold. Firstly, higher
order representations although essential to obtain a reasonable hyperon
spectrum appear with a quite small weight in the low-lying hyperon wavefunction.
Secondly, the collective operators that contain stronger ``non-octet" contributions
(as i.e.  $R_{8,8} R_{{\pi^-},a} R_{{K^+},a}$ ) appear in terms 
proportional to $\phi_{SB}$ which is, numerically, one order of magnitude smaller
than the leading contributions (terms proportional to $\phi_{SK}$). 
Nevertheless, as already mentioned above, these ``dynamical" symmetry breaking
terms are important to obtain good mass splittings. 
Also in Table~1 we observe that the other calculated ratios are (in absolute
value) somewhat larger than the empirical ones. However, they are basically
of the same quality as those of the PTA or the BSA. 
In any case, the main success of the present model over the other soliton
approaches is in the prediction of the absolute values of the decay
amplitudes. Since we have seen that the ratios to the
$A(\Lambda^0_-)$ amplitude are reasonably described, it is enough
to consider the absolute value of such quantity. From Table~2 we see
that our calculated value is in good agreement with the empirical one.
The improvement with respect to the PTA and BSA is very significant
and shows that the
use of empirical values for the model parameters is essential to
describe the weak decay amplitudes correctly. In Table~2 we also see that our 
prediction is somewhat better than that of the QM. However, it should be noticed
that in the QM this amplitude is particularly problematic. In general, the QM 
results are of the same quality than ours. 

Finally, we discuss the role of strong interaction enhancement factors
used in previous soliton calculations. These factors were introduced
in the context of the quark model to account for hard gluon 
exchanges\cite{SVZ77}. It is not clear whether they should 
also be used in soliton calculations since, in principle, they
could be already contained in the non-perturbative soliton
currents. This question was already raised in the context of
the $SU(2)$ soliton model (see i.e. Ref.\cite{Mei92}). The results
corresponding to PTA and BSA given in Table~2 do include an enhancement
factor $c_1 \approx 2.6$. In our calculation we have not used
such factor. The good agreement we have found with respect to the empirical
value of $A(\Lambda^0_-)$ seems to clearly indicate that there is
no need for the enhancement factors in the soliton models.

In conclusion, we have studied the S-wave non-leptonic weak decay amplitudes
of the hyperon in the context of an $SU(3)$ soliton model in which
strangeness degrees of freedom are introduced through collective
variables and symmetry breaking terms are diagonalized exactly.
We have obtained a very nice agreement of both the relative and absolute
amplitudes with the corresponding empirical values. In fact, we have
found a substantial improvement in the prediction of the absolute amplitudes with
respect to previous soliton calculations. Finally, we have seen
that although the present approach includes some configuration mixing
the corresponding impact on the ``octet dominance" is small enough
to guarantee that the empirical ``$\Delta I = 1/2$" is still
well satisfied.

%\vspace{1.cm}

\vskip 3cm

%\pagebreak
 
%--------------------- End Table 1 ----------------------------------
\begin{table}[h]
\begin{center}
Table 1: The nonleptonic hyperon decay amplitudes taken with respect to $A(\Lambda_-^0)$
amplitude.

\vspace{1.cm}

\begin{tabular}{|c|c|c|c|c|}\hline
                  & This work & PTA \cite{PT85} & BSA \cite{KSO90} 
& Empirical \cite{DGH86}\\ \hline
$A(\Sigma_+^+) $  & $\ $ 0.05     &  $\ $   0.00  &   $\ $ 0.00  &  $\ $    0.04    \\ \hline
$A(\Sigma_0^+) $  & --1.26     &   --1.00  &  --1.00  &    --1.00    \\ \hline
$A(\Sigma_-^-) $  &  $\ $ 1.74     &  $\ $   1.43  &  $\ $  1.41  &   $\ $   1.31    \\ \hline
$A(\Xi_-^-)    $  & --1.54     &   --1.43  &  --1.73  &    --1.39    \\ \hline
\end{tabular}
\end{center}
\end{table}
%--------------------- End Table 1 ----------------------------------

%--------------------- End Table 2 ----------------------------------
\begin{table}[h]
\begin{center}
Table 2: Absolute value of the S-wave $\Lambda \to p \pi^-$ decay amplitude.

\vspace{1.cm}

\begin{tabular}{|c|c|}\hline
     & $A(\Lambda_-^0)$  $\times 10^6$ \\ \hline
This work              & 0.35 \\ \hline 
PTA \cite{PT85}        & 0.07 \\ \hline
BSA \cite{KSO90}       & 0.08 \\ \hline
QM\cite{TT81}         & 0.21 \\ \hline
Empirical\cite{DGH86}  & 0.32 \\ \hline
\end{tabular}
\end{center}
\end{table}
%--------------------- End Table 1 ----------------------------------
\end{document}